\documentstyle[12pt]{article}
\normalbaselines
\begin{document}
\bibliographystyle{unsrt}

\vbox{\vspace{6mm}}

\begin{center}
{\large \bf SQUEEZING IN A 2-D GENERALIZED OSCILLATOR} \\ [7mm]
Octavio Casta\~nos and Ram\'on L\'opez-Pe\~na \\
{\it Instituto de Ciencias Nucleares, UNAM \\
Circuito Exterior, C.U., Apdo. Postal 70-543 \\
04510 M\'exico, D. F., M\'exico} \\ [5mm]
Vladimir I. Man'ko \\
{\it Physical Institute of the Russian Republic \\
Leninsky Prospect 53. Moscow, Russia} \\ [5mm]
\end{center}

\vspace{2mm}

\begin{abstract}
A two-dimensional generalized oscillator with time-dependent
parameters is considered to study the two-mode squeezing phenomena.
Specific choices of the parameters are used to determine the
dispersion matrix and analytic expressions, in terms of standard
hermite polynomials, of the wavefunctions and photon distributions.
\end{abstract}

\section{Introduction}
In the middle of the sixties and beginning of the seventies a set of quantum
states of the electromagnetic field were observed which have less uncertainty
in one quadrature than a coherent state [1-3].
These one-mode squeezed states have generated big
expectations in optical communication systems [4].  In some quantized fields,
the interaction hamiltonians occur only between pairs of modes and then to
understand the main features of the system, one restricts to study one and two
normal modes .  In the last
decade two-mode squeezing phenomena have attracted attention to
study properties of noise and correlations [5-8].  Recently the accidental
degeneracy of a two-dimensional (2-D) harmonic oscillator with frequency
$\omega_0$ plus an
interaction proportional to the z-th projection of the angular
momentum was studied~[9].  This system was called the
generalized 2-D harmonic oscillator because presents a
bigger accidental degeneracy depending on the strength $\lambda$
of the angular momentum interaction.  This model was generalized [10]
to include time-dependent parameters, $m = m_0 f (t)$ and $\lambda =
\omega _0 \lambda_0(t)$.  If we take $f(0) = \lambda_0(0) = 1$;  the
hamiltonian, for $t = 0$, represents a charged particle moving in a constant
magnetic field .

The aim of this work is to study two-mode squeezing phenomena with
this model because it demonstrates the change of dispersions
due to variation of the mass and coupling constant during the
evolution.  In the framework of quantum optics the hamiltonian is built by:
the operator $(1/f + f)\vec a^\dagger \cdot \vec a$, that causes a
time-dependent exchange of kinetic and potential energies within each
mode; the interaction $1/2 (1/f -f)(\vec a^\dagger \cdot \vec
a^\dagger + \vec a\cdot \vec a)$, which describes a degenerate
two-photon interaction; and the potential $i \lambda_0 (t) (a^\dagger
_2 a_1 -a^\dagger_1 a_2)$, that is a mode mixing operator.

The solution of the corresponding time dependent Schroedinger equation is
obtained through the theory of integrals of
motion~[11].  By means of Noether's
theorem, using a special variation [10], we construct the linear time
dependent integrals of the motion.  The resulting quantum invariants
are given in terms of the positions and momenta operators [10,11] by
	\begin{equation}
		\vec P(t) = \lambda_1 \vec p + \lambda_2 \vec q,
		\quad \vec Q (t) = \lambda_3 \vec p + \lambda_4
		\vec q \, ,
	\end{equation}
with the initial conditions $\vec P (0) = \vec p$ and $\vec Q(0) =
\vec q$, so that the $2\times 2$ matrices previously introduced
satisfy $\lambda_1(0) = \lambda_4(0) = I_2$ and $\lambda_3(0) =
\lambda_2(0) = 0$.  The operators $\vec A(t) = 1
/\sqrt{2} \, [\vec Q(t)/l + il/\hbar \vec P(t)]$ and its hermitean
conjugate, can be constructed with the matrices
	\begin{equation}
		\lambda_p = {1\over l} \lambda_3 + {il \over \hbar}
		\lambda_1, \quad \lambda_q = {1\over l} \lambda_4 +
		{il\over \hbar} \lambda_2 \,  ,
	\end{equation}
with $l = \sqrt{{\hbar \over m_0 \omega_0}}$ defining the
oscillator length.  These integrals of motion also are given in terms of the
creation and annihilation photon operators
	\begin{equation}
		\vec A(t) = M_1 \vec a + M_2 \vec a^\dagger, \quad
		\vec A^\dagger (t) = \quad M_3 \vec a + M_4 \vec
		a^\dagger \,  .
	\end{equation}
With the initial conditions $\vec A(0) = \vec a$ and $\vec A^\dagger
(0) = \vec a^\dagger$, the matrices defined in (3)
comply with $M_1(0) = M_4(0) = I_2$ and $M_3(0) = M_2(0) = 0$.  The
$\lambda_k$'s, $M_k$'s, $\lambda_p$ and $\lambda_q$ are entries of
symplectic matrices in four dimensions because the invariants (1) and
(3) satisfy the commutation relations of Heisenberg-Weyl algebras.

In the present work, we study the behavior of the model for
$\lambda_0(t)$ an arbitrary function of time and considering two
kinds of varying masses, {\it i.e.}, two choices for the function
$f(t)$, namely:
	\begin{equation}
		f(t) = \hbox{exp}(\gamma t) \ \ ;
	\end{equation}
	\begin{equation}
		f(t) = \cases{1\hfill \ \ , & $ t \leq 0 $ \cr
		\cosh^2\Omega_0 t \hfill \ \ , & $ 0 \leq t \leq T$ \cr
		\{\Omega_0 (t -T) \sinh \Omega_0 T + \cosh \Omega_0
		T\}^2 \hfill \ \ , & $T\leq t$ \cr} \ \ .
	\end{equation}
For these two cases the $\lambda_k$ matrices take the general form
	\begin{equation}
		\lambda_k = \mu_k \ {\bf R} = \mu_k \left(
		\matrix{\hfill \cos\theta & \sin\theta \cr
		\hfill - \sin\theta & \cos\theta \cr}\right); \ \ k = 1, \,
		2, \, 3, \, 4;
	\end{equation}
where the definition $
 \theta = \int^t_0 \omega_0
\lambda_0(\tau)d\tau$ was used.  The analytic expressions for the $\mu_k$'s
functions are given in Ref. [10].  In the
next sections we determine the coherent and Fock-like states, the
photon distributions and the dispersion matrices in terms of these
$\mu_k$'s.

\section{Squeezed Coherent and Fock States}
The coherent-like states are obtained by solving the differential
equation $\vec  A (t) \Phi_0 (\vec q, t) = 0$ with $\vec A (t)$ given
in Eq. (3).  This solution yields the vacuum state of the physical
system, and its phase is chosen to guarantee that satisfies the time dependent
Schroedinger equation. The expression for the ground
state wavefunction is
	\begin{equation}
		\Phi_0 (\vec q, t) = { 1\over \sqrt{2\pi
		\hbar^2}\mu_p} \ \hbox{exp} \biggl \{- {i\over
		2\hbar} {\mu_q \over \mu_p} \vec q \cdot \vec
		q\biggr\} \ \ .
	\end{equation}
To get the last expression the relation (2) was used and the functions
$\mu_p = {1 \over \sqrt{2}} \left( {il \over \hbar} \mu_1 + {1\over
l} \mu_3\right)$ and $\mu_q = {1\over \sqrt{2}} \left({il \over
\hbar} \mu_2 + {1\over l}\mu_4\right)$ were defined. To obtain the general
expression for the eigenstates in the
coordinate representation one needs to apply the unitary operator
$\hat D (\alpha) = \hbox{exp}\{\vec \alpha \cdot \vec A^\dagger -
\vec \alpha^\ast \cdot \vec A\}$, which is an invariant, to the
vacuum wavefunction (7), {\it i.e.},
	\begin{equation}
		\Phi_\alpha(\vec q, t) = \ \hbox{exp}\biggl\{ -
		{|\alpha|^2 \over 2} + {1 \over 2} {\mu^\ast_p
		\over \mu_p} \, \vec \alpha \cdot \vec\alpha + {i \over
		\hbar \mu_p} \vec{q} \ \tilde{\bf R} \vec \alpha\biggr\}
		\Phi_0 (\vec q, t) \ \ .
	\end{equation}
These are expressed in terms of multi-dimensional
Hermite polynomials [12] through the relation
	\begin{equation}
		\hbox{exp} \biggl( - {1\over 2} \, u \, \vec \alpha^\ast
		\cdot \vec \alpha^\ast + v \, \vec \alpha^\ast \, \tilde
		{\bf R} \, \vec\gamma \biggl) = \sum^\infty_{n_1, n_2
		=0} {\alpha^{\ast n_1}_1 \over n_1!} {\alpha^{\ast
		n_2} \over n_2!} {\bf H}^{\{ u{\bf I}_2\}}_{n_1, n_2}
		\biggl( {v \over u} \tilde{\bf R} \vec \gamma\biggr)
		\  \ .
	\end{equation}
Substituting the last expression into (8) and using the form of the
coherent-like states in the Fock-like representation, we get the Fock-like
eigenstates in the coordinate representation:
	\begin{equation}
		\langle \vec q | n_1 n_2 \rangle = \Phi_0 ( \vec q, t)
		\ {\bf H}^{ \biggl\{ - {\mu^\ast_p \over \mu_p} {\bf
		I_2} \biggr\} }_{n_1, n_2} \biggl( - {i \over \hbar
		\mu^\ast_p} {\bf R} \vec q \biggr) \ .
	\end{equation}
These multi-dimensional Hermite polynomials are rewritten as a product of two
standard one-dimensional Hermite polynomials [12] as follows:
\begin{eqnarray}
	{\bf H}^{ \biggl\{ { \mu^\ast_p \over
		\mu_p} {\bf I_2} \biggr\} }_{n_1, n_2} \biggl( - {i
		\over \hbar \mu^\ast_p} {\bf R} \vec q \biggr) & = &
		\biggl( - {\mu^\ast_p \over 2 \mu_p} \biggr)^{(n_1 +
		n_2 ) / 2} H_{n_1} \biggl( {1 \over \sqrt{2} \hbar | \mu_p
		| } [ \cos \theta \ q_1 + \sin \theta \ q_2 ] \biggr)
		\nonumber \\
		 &  & \mbox{ } \times H_{n_2} \biggl( {1 \over \sqrt{2}
		\hbar| \mu_p | } [ - \sin \theta \ q_1 + \cos \theta \
		q_2 ] \biggr) \ ,
\end{eqnarray}
where we use the explicit expression of matrix ${\bf R}$.  These Fock
(10) and coherent (8) -like states represent squeezed and correlated
eigenstates of the system as it will be shown further.

\section{Propagator}
The propagator in the coherent state representation is given by the
matrix elements of the evolution operator $U(t)$, which will be
obtained by means of the theory of time dependent integrals of motion~[11]. If
$\vec I(t)$ is an integral of motion then satisfies $\vec I(t)
\hat U(t) = \hat U(t) \vec I(0) $.  Taking its matrix elements with
respect to the coherent states, we get a linear system of
differential equations, which can be solved.  Thus the propagator
takes the form
	\begin{equation}
		G(\vec\alpha^\ast, \vec\gamma, t) = {\hbox{exp} \ ( -
		|\vec \alpha|^2/2 - |\vec\gamma|^2/2) \over \sqrt{det
		M_1}} \hbox{exp} \biggl( - {1\over 2} \vec\alpha^\ast
		M^{-1}_1 M_2 \vec\alpha^\ast + \vec\alpha^\ast
		M_1^{-1} \vec \gamma + {1\over 2} \vec\gamma M_3
		M_1^{-1} \vec\gamma\biggr) \ \ .
	\end{equation}

For the cases (4) and (5), the following relations are satisfied
	\begin{equation}
		\sqrt{det M_1} = {1\over \sqrt{2}} \biggl( l\mu_q -
		{i\hbar \over l} \mu_p \biggr) \equiv g_1, \quad
		M^{-1}_1 M_2 = {1\over \sqrt{2} g_1} \left( l \mu_q +
		{i\hbar \over l} \mu_p\right) {\bf I_2} \equiv g_2
		{\bf I_2} \ \ ,
	\end{equation}
	\begin{equation}
		M^{-1}_1 = {1\over g_1} \tilde {\bf R} \ , \quad M_3
		M^{-1}_1 = {1\over \sqrt{2}g_1} \biggl(l \mu^\ast_q +
		{i\hbar \over l} \mu^\ast_p\biggr) I_2 \equiv g_4
		{\bf I_2} \ \ .
	\end{equation}
Substituting these relations into the Eq.(12) we get the propagator,
which through Eq. (9) can be expressed in terms of multi-dimensional
Hermite polynomials.  If we compare with the power series expansion
of the propagator we get the probability amplitude for having $n_1$
and $n_2$ photons in the coherent-like state $|\vec\gamma, t \rangle$, {\it
i.e.},
	\begin{equation}
		\langle n_1 n_2 | \vec\gamma, t \rangle = {1 \over
		g_1 \sqrt{n_1 !N_2!}} \hbox{exp} \ \biggl( -
		{|\vec\gamma|^2 \over 2} + {1\over 2} g_4 \vec \gamma
		\cdot \vec\gamma \biggr) {\bf H}^{\{g_2 {\bf
		I_2}\}}_{n_1, n_2} \biggl( {1\over g_1 g_2} \tilde
		{\bf R} \vec\gamma \biggr) \ \ .
	\end{equation}

By means of the Eq.(12) this amplitude can be
rewritten in terms of standard Hermite polynomials [12].  The squared
absolute value of this amplitude yields the photon distribution
function of the system, $W_{n_1 n_2} (\vec\gamma, t) = |\langle n_1
n_2|\vec\gamma, t\rangle|^2$.  This will let us calculate, at least
formally, the mean, $\langle N_k \rangle$, and the mean squared
fluctuation of the number of photons, $(\Delta N_k)^2$, in direction $k$, which
are present in the coherent state $|\vec\gamma, t\rangle$. The expectation
values of $N_k$ and $N^2_k$ are evaluated directly using the expressions of the
creation and annihilation photon operators in terms of the integrals of the
motion (3), and the commutation properties for these invariants. For the vacuum
state one has
	\begin{equation}
		\langle N_k \rangle = {1\over 4} \biggl\{ (\mu_1 -
		\mu_4)^2 + \left ( m_0 \omega_0 \mu_3 + {1\over m_0
		\omega_0} \mu_2 \right)^2 \biggr\} \, ,
	\end{equation}
\begin{eqnarray}
		\langle N^2_k \rangle & = & {
		1\over 8} \biggl\{ (\mu_1 - \mu_4)^2 + \bigl( m_0
		\omega_0 \mu_3 + {1\over m_0
		\omega_0} \mu_2 \bigr)^2\biggr\} \nonumber \\
		& & \biggl\{ (\mu_1 +
		\mu_4)^2 + \bigl( m_0 \omega_0 \mu_3 -
		{1\over m_0 \omega_0} \mu_2
		\bigr)^2\biggr\}
		 \mbox{  } +  {1 \over 16}  \biggl\{
		(\mu_1 -
\mu_4)^2 + \bigl( m_0 \omega_0  \mu_3 +
		{1\over m_0 \omega_0} \mu_2\bigr)^2
		\biggr\}^2 \, .
\end{eqnarray}
With these expressions, we evaluate the ratio of the mean
squared fluctuation $(\Delta N_k)^2$ and the mean number of photons
$\langle N_k\rangle$, which determines the nature of the distribution function
of the system:
	\begin{equation}
		{(\Delta N_k)^2 \over \langle N_k \rangle} = {1\over
		2} \biggl\{ (\mu_1 + \mu_4)^2 + \biggl( m_0\omega_0
		\mu_3 - {1\over m_0 \omega_0} \mu_2\biggr) ^2
		\biggr\}
 \ \ .
	\end{equation}
For the cases (4) and (5) the ratio is greater than one  when $t > 0$, which
implies that we have a super-Poissonian photon distribution function. For
$t=0$, there is a discontinuity in the ratio, which is obtained by comparing
the following limiting procedures: making $t \rightarrow 0$ and then
$\vec{\alpha} \rightarrow 0$, and conversely.

\section{Dispersion Matrices}
The dispersion matrix can be written in terms of $2\times 2$ matrices
characterizing the dispersions in the positions and momenta operators
and the correlation between them.  Besides for the cases under study,
due to (6), they take the form
	\begin{equation}
		\sigma^2_{pp}(t) = {1\over 2} \hbar m_0\omega_0
		\biggl( {1\over (m_0 \omega_0)^2 }\mu^2_2 +
		\mu^2_4\biggr)  \ {\bf I_2} \ \ ,
	\end{equation}
	\begin{equation}
		\sigma^2_{qp}(t) = - {\hbar \over 2} \biggl( {1 \over
		m_0\omega_0} \mu_1\mu_2 + m_0\omega_0
		\mu_3\mu_4\biggr) \ {\bf I_2} \ \ ,
	\end{equation}
	\begin{equation}
		\sigma^2_{qq}(t) = {1\over 2} {\hbar\over
		m_0\omega_0} \bigl(\mu^2_1 +
		(m_0\omega_0)^2\mu^2_3\bigr) \ {\bf I_2} \ \ .
	\end{equation}
The corresponding correlation matrices for the creation
and annihilation operators are obtained immediately from the last expressions;
they are given by
	\begin{equation}
		\sigma^2_{aa} = {1\over 4} \biggl\{ \mu^2_1 - \mu^2_4
		+ (m_0\omega_0)^2 \mu^2_3 - {\mu^2_2 \over
		(m_0\omega_0)^2} - 2i \biggl( {1\over m_0\omega_0}
		\mu_1\mu_2 + m_0\omega_0 \mu_3\mu_4
		\biggr)\biggr\} \
\ ,
	\end{equation}
	\begin{equation}
		\sigma^2_{a^\dagger a} = {1\over 4} \biggl( \mu^2_1 +
		\mu^2_4 + (m_0\omega_0)^2 \mu^2_3 + {\mu^2_2 \over
		(m_0 \omega_0)^2}
{m_0\omega_0 \over \hbar} 		\biggr)\\ ,
	\end{equation}
	\begin{equation}
		\sigma^2_{a^\dagger a^\dagger} = {1\over 4} \biggl\{
		\mu^2_1 - \mu^2_4 + (m_0 \omega_0)^2 \mu^2_3 -
		{\mu^2_2 \over (m_0 \omega_0)^2} + 2i \biggl( {1\over
		m_0\omega_0} \mu_1\mu_2 + m_0\omega_0
		\mu_3\mu_4\biggr) \biggr\} \ \ .
	\end{equation}

\vspace{13truecm}
\begin{quotation}
Fig. 1. Dispersion and correlation matrices behavior in positions and momenta
space for the studied cases  in this paper: (a) corresponds to Eq.(4), and (b),
to Eq.(5).
\end{quotation}

The behavior of the dispersion matrices is illustrated
in Fig. 1.  For the case (4), we choose the parameters $\gamma = 0.1$ and $m_0
=
\omega_0 =1$.  It is seen that there is squeezing for the
coordinates and stretching for the momenta.  Also one notes that $\sigma_{pq}$
is a negative function and therefore there
are one-mode correlations between the coordinates and the momenta.
If we reverse the sign of $\gamma$, the roles between the dispersion
for coordinates and momenta are interchanged, and $\sigma_{pq}$
becomes positive.
In the case (5), we use the parameters
$\Omega_0 = 0.15, \ T = 10$, and $m_0 = \omega_0 = 1$. In spite of
the mass is
different that in the previous example, the general trends are
similar.  For example, the $\sigma_{pp}$ is an increasing function of
time starting from its minimum value at $t \leq 0$, and there is
squeezing for the $\sigma_{qq}$.  The main difference appears in the
correlation $\sigma_{pq}$: in this case, it can be positive for large times,
while in
the previous one is negative or zero for any time.

 \section{Acknowledgments}
Work supported in part by projects UNAM-DGAPA IN-103091 and CONACYT
1570-E9208.

\end{document}